# Increasing risk of oppressive heatwaves over India in the future warming

Naveen Sudharsan[a], Jitendra Singh[b], Subimal Ghosh[c,d], Subhankar Karmakar[a,d,e]∗

[a] Environmental Science and Engineering Department, Indian Institute of Technology Bombay, Maharashtra, India

[b] Department of Environmental Systems Science, ETH Zürich, Zürich, Switzerland

[c] Department of Civil Engineering, Indian Institute of Technology Bombay, Maharashtra, India

[d] Interdisciplinary Program in Climate Studies, Indian Institute of Technology Bombay, Maharashtra, India

[e] Centre for Urban Science & Engineering, Indian Institute of Technology Bombay, Maharashtra, India

## Abstract

The frequency of heatwave events has been increasing with climate change across the globe, causing devastating impacts on human and natural systems. Particularly, the co-occurrence of high temperature with high humidity has the potential of worse impacts on human health than extreme temperature alone. We use observations to examine the characteristics of extreme (high temperature and low humidity) and oppressive (high temperature and high humidity) heatwaves and their impacts on human mortality. We also use climate model simulations to understand the likelihood of heatwaves under varied warming conditions in future climates. We find an increasing trend in the number of days of extreme and oppressive heatwaves, with surprisingly a higher rate of increase in oppressive heatwaves. We also observe a higher correlation between oppressive heatwaves and historical heat-stress-related human deaths than extreme heatwaves, indicating more substantial impacts of oppressive heatwaves on the human population than extreme heatwaves. Further, our results show a five-fold increase in the number of days of oppressive heatwaves under 1.5° warming in both mid (2035-2065) and end (2070-2100) of the 21st century, relative to the historical period (1975 to 2005) whereas the number of extreme heatwave days remains relatively constant in mid and end of this century. Remarkably, 2° warming results in an eight-fold and almost two-fold increase in the number of days of oppressive heatwaves by the end of the century relative to the historical period and 1.5° warming conditions. These results suggest that limiting the mean global warming to 1.5° can reduce the likelihood of oppressive and extreme heatwaves by 44% and 25% by the end of the century relative to the 2° warming world. Moreover, the remarkable increase in oppressive heatwave days highlights the elevated risk of heatwaves over densely populated countries and indicates an imminent need for adaptation measures.

**Introduction**

Anomalously warm conditions, referred to as heatwaves, are often associated with high socioecological risks (Kornhuber et al., 2020; Saeed et al., 2021). Several multi-country (Calleja-Agius et al., 2021; Campbell et al., 2018; Hajat et al., 2006; Jay & Capon, 2018; McMichael et al., 2006; Xu et al., 2020) and regional (Cheng et al., 2018; Lissner et al., 2012; Lowe et al., 2011; Mazdiyasni et al., 2017; Murari et al., 2015; Sun et al., 2014; Yang et al., 2019) studies have highlighted the severe impacts of heatwaves on human health including human mortality and morbidity. Although heatwaves impact the human population regardless of their age and health conditions, they pose a high risk to the population with existing medical conditions, relatively older population, and physical labor working outside for a long time (Calleja-Agius et al., 2021; Perkins et al., 2012a). For instance, the European heatwaves in 2003 killed more than 40000 people across Europe (Beniston, 2004; García-Herrera et al., 2010; Robine et al., 2008). Similarly, the Russian heatwave in 2010 killed more than 55000 people, which is considered the deadliest heatwave that humankind has ever experienced (Barriopedro et al., 2011; Dole et al., 2011; Otto et al., 2012). Likewise, India has experienced the death of thousands of people due to multiple heatwaves that occurred in 1988, 1998, and 2015 (De, U. S., Dube & Rao, 2005; De & Mukhopadhyay, 1998; Guleria, 2018; Jenamani, 2012; Satyanarayana & Rao, 2020).

Moreover, heatwaves present many indirect negative impacts on physical environment systems (Horton et al., 2016). An increase in temperature elevates the evapotranspiration demand could result in drought conditions (Sharma & Mujumdar, 2017), and thereby may cause crop failures and turn pastureland into barren land (Teuling et al., 2013). Such impacts can have ripple effects on food security and pastureland required for grazing livestock (Kogan et al., 2019; Miyan, 2015). Since 1980, the increasing rate of global mean temperature is nearly 3-fold (0.18°C/decade) than that since 1880 (0.07°C/decade) (Lindsey & Dahlman, 2020), suggesting that the world is moving towards a warmer one. Such future warming can elevate the frequency and intensity of heatwaves and their cascading impacts on human and natural systems across the globe (Fischer & Knutti, 2013; Meehl & Tebaldi, 2004; Murari et al., 2015). In particular, South Asia, a habitant of one-fifth of the global population, is at a greater risk of heatwaves in warmer future climates (IPCC, 2013; Saeed et al., 2021).

Although heatwaves are defined as consecutive days of anomalous temperature, the heatwaves definition differs according to geographical locations and temporal/spatial scales of studies (Hajat et al., 2006; Khan et al., 2018; Mazdiyasni et al., 2017; J. Nairn et al., 2009; Rohini et al., 2016). For example, the India Meteorological Department (IMD), the principal agency responsible for meteorological observations in India, identifies heatwaves as three or more days with a temperature exceeding a predetermined threshold temperature based on topography (i.e., above 45°C in plains and above 40°C in hilly areas). Instead of considering such a fixed threshold for all locations with the same topography, several studies employ a quantile threshold approach, which is considered a robust approach yet region-specific to understanding the heatwave effect (Perkins et al., 2012b). The percentile-based approach developed by the Expert Team on Climate Change Detection and Indices (ETCCDI) defines a heatwave event as six or more consecutive days with maximum daytime temperature exceeding the 90th percentile for the long-term historical period (Alexander et al., 2006). However, the Effective Heat Factor (EHF) approach introduced by Nairn et al. (2009) employs both daily maximum temperature ($T_{max}$) and daily minimum temperature ($T_{min}$) to identify heatwaves, which makes it suitable for human health applications as it indirectly accounts for daytime and nighttime temperatures (Perkins and Alexander, 2012; Ford and Schoof, 2017). This EHF-based approach is widely used and also called the universal heatwave index (Varghese et al., 2019).

The severity of heatwaves can be attributed to a single factor, like extreme temperature, or to a combination of multiple climatic variables, such as temperature and humidity, where not all variables need to be anomalously high (Leonard et al., 2014). Humidity is as vital as the surface temperature in heatwave analysis, as it affects the heat stress experienced by humans (Kang & Eltahir, 2018). An increase in humidity adversely affects the rate of evaporative cooling, failing the thermoregulation process in organisms and leading to an increase in core body temperature (Kravchenko et al., 2013). The humidity is relatively high in tropical regions compared to subtropical and polar regions. Thus, the estimation of heatwave based on temperature alone may underestimate the heatwave characteristics and their impacts on human health as high humidity along with sweltering days contributes to amplifying the heat stress due to heatwaves (Haldane 1905, Brunt 1943, Thom 1959; Conti et al., 2005; Dematte et al., 1998; Fischer & Knutti, 2013). Despite that the simultaneous occurrence of high temperature with high humidity causes severe impacts on human health, only a few past studies have accounted the effect of humidity in heatwave estimations (Fischer & Knutti 2012, Smith et al. 2016, Mazdiyasni et al. 2017, Bishop-williams et al. 2015, Mora et al. 2017, Russo et al. 2017, Buzan & Huber 2020).

Extreme temperature-related deaths are often classified incorrectly and seldom recognized as being deaths directly caused due to extreme temperatures. Instead, these are usually attributed to underlying medical conditions such as cardiovascular or respiratory diseases, which indeed contribute to increased vulnerability to extreme temperatures (Calleja-Agius et al., 2021). Thermoregulation is the process by which a body effectively handles its thermal stress, particularly in healthy individuals with no pre-existing morbid conditions. In contrast, the thermoregulation process is less effective in physically weaker groups, such as the geriatric population (significantly in age groups above 65), kids, and people with pre-existing medical conditions (like cardiovascular or respiratory diseases) (Russo et al., 2017). The risk of mortality/morbidity to heatwaves is exceptionally high for these groups and further amplifies in humid conditions. Additionally, the simultaneous occurrence of high heat and high humidity poses a significant risk to individuals who work long hours in daylight.

The present study examines the historical changes in the heatwave characteristics and their association with human mortality in India. Given the importance of humidity in heatwave estimation (Wehner et al., 2016) and India experiences high humidity conditions due to its geographical location (Parishwad et al., 1998), we classify heatwaves into oppressive (high temperature and high humidity) and extreme (high temperature and low humidity) heatwaves. We further examine the likelihood of heatwave events under future 1.5° and 2° warming conditions relative to the pre-industrial period using daily maximum temperature, daily mean temperature, and specific humidity simulations from the Community Earth System Model Large Ensemble Numerical Simulation (CESM-LENS).

**Data and Methodology**

We use the India Meteorological Department (IMD) gridded daily temperature data at 1° spatial resolution and National Oceanic and Atmospheric Administration (NOAA) - National Centre for Environmental Prediction (NCEP) /National Center for Atmospheric Research (NCAR) reanalysis 1 specific humidity at 2.5° spatial resolution (interpolated to 1°) from 1951 to 2013. We also used a five-member large ensemble from the CESM-LE simulations and bias-corrected the model output following the methods provided by Salvi et al. (2013). CESM is a fully coupled, community, global climate model that provides the Earth's past, present, and future climate. CESM-LE daily outputs are recognized as suitable for identifying mechanisms in future heatwave projections and the effect of internal climate variations on heat stress metrics (Kay et al., 2015). We project the oppressive heatwave days and extreme heatwave days for the near future (2035 to 2065) and the far future (2070 to 2100) to understand the changes in heatwaves during these two time periods relative to historical climate (1975 to 2005).

We use the excess heat factor (EHF) approach to characterize the oppressive and extreme heatwave events (Ford and Schoof, 2017; Langlois et al., 2013; Nairn et al., 2009; Nairn & Ostendorf, 2018; Perkins et al., 2012; Rohini et al., 2016; Wang et al., 2018). EHF (Equation 1) represents the amplification of the long-term temperature anomaly ($EHI_{accl}$) by the short-term temperature anomaly ($EHI_{sig}$) (Langlois et al., 2013). It expresses the impact of the increase in both daytime and nighttime temperature, thus making it the suitable method to evaluate the effect of heatwaves on human health compared to other indices, which use only maximum or minimum daily temperature (Nairn & Ostendorf, 2018; Nairn & Fawcett, 2015).

$$EHF = EHI_{sig} \times \max(1, EHI_{accl}) \quad (1)$$

$$EHI_{sig} = \frac{T_{i-2} + T_{i-1} + T_i}{3} - T_{85} \quad (2)$$

$$EHI_{accl} = \frac{T_{i-2} + T_{i-1} + T_i}{3} - \frac{T_{i-32} + T_{i-31} + \cdots + T_{i-3}}{30} \quad (3)$$

$$T_E = T_{avg} + \frac{L_v q}{C_p} \quad (4)$$

where, $T_{85}$ is the 85$^{th}$ percentile temperature (°C) of the hottest month (Mazdiyasni et al., 2017), $T_i$ is the temperature ($T_{max}$ or $T_E$, °C) for the i$^{th}$ day. $T_{avg}$ is the daily average temperature (°C), q is specific humidity (Kg/Kg), $L_v$ is the latent heat of vaporization (J/Kg), and $C_p$ is the specific heat at constant pressure (J/Kg°C) (Schoof et al., 2015). Excess heat index significance ($EHI_{sig}$) (Equation 2) denotes the anomalous high temperature that is not adequately cooled overnight due to high nighttime temperature, and excess heat index acclimatization ($EHI_{accl}$) (Equation 3) signifies the period, which is warmer than the recent past. Here we consider thirty days as the recent past as the human body takes at least two weeks to get acclimatized to temperature change (Nairn & Fawcett, 2015). The framework characterizes the heatwaves by considering EHF based on $T_{max}$ and daily equivalent temperature ($T_E$) (Equation 4), such that an oppressive heatwave condition befalls when both factors are positive. Whereas when $T_{max}$ based EHF is positive and $T_E$ based EHF is negative, it denotes an extreme heatwave condition. Additionally, we evaluate the effect of a sudden increase in temperature by considering that only an increase of more than 1°C from the recent past affects the population adversely. Therefore, we consider the day to be a heatwave day only if the $EHI_{accl}$ is more than 1°C. Further, we examine the association of heat wave days with Morality in India from 1976 to 2006 using the heatwave mortality data in IMD annual reports (IMD, 2010).

At the 21st Conference of the Parties in Paris (2015), the participating nations pledged to maintain the temperature below 2°C above the pre-industrial levels while attempting to keep it below 1.5°C (Council, 2015). Thus, we use CESM-LE simulations with 1.5 °C and 2 °C warming scenarios (Sanderson et al., 2017) from 2006 to 2100 to understand the repercussions of the projected temperature increase on heatwave characteristics. Further, we estimate the changes in the future heatwave characteristics relative to the historical period considered from 1975 to 2005.

**Results and Discussions**

*Historical trends in heatwave characteristics in the observations*

Historically, we find that northwestern, central, and eastern parts of India experienced >15 heatwave events with a one-day duration from 1953 to 2013 (Figure 1a). However, the frequency of heatwave events with two to five days is nearly 2-fold (~>25 events) relative to events with one-day duration over the same regions and period (Figure 1b), consistent with Ratnam et al. (2016) and Rohini et al. (2016). We further classify these heatwave events into oppressive and extreme heatwave events (see *methods* for details). The number of extreme heatwaves with shorter (one day) and longer (two to five days and more than five days) duration is almost double that of the oppressive heatwaves across the same regions (Figure 1d-i). However, the total number of oppressive heatwave days increased over a higher land fraction of India (~23% area) compared to the area (~17%) where extreme heatwave days increased (Figure 2). Aggregately, the number of oppressive heatwave days significantly (at a 5% significance level) has increased in India in the last six decades, whereas the number of extreme heatwave days remains unchanged (Figure S2a-b). Similarly, total (oppressive + extreme) heatwave days increase significantly over India (Figure S2c). Such contrasting trends in oppressive and extreme heatwave days highlight the potential of experiencing a higher number of oppressive heatwave days than extreme heatwave days in the future. Further, we examine the trends in heatwave days across seven meteorologically homogeneous zones (southern zone, central zone, western zone, northern zone, northeastern zone, northeast hilly zone, and Jammu & Kashmir zone) across India (Figure S7; reference to the zone classification) (Figure S3) to understand the spatial pattern in heatwaves day trends. We find that all regions except Northern and Jammu & Kashmir Zones show a significant (at 5% significance level) increasing trend in oppressive heatwave days in the last decades. In contrast, only the western zone shows a significant increasing trend in extreme heatwave days (Figure S3).

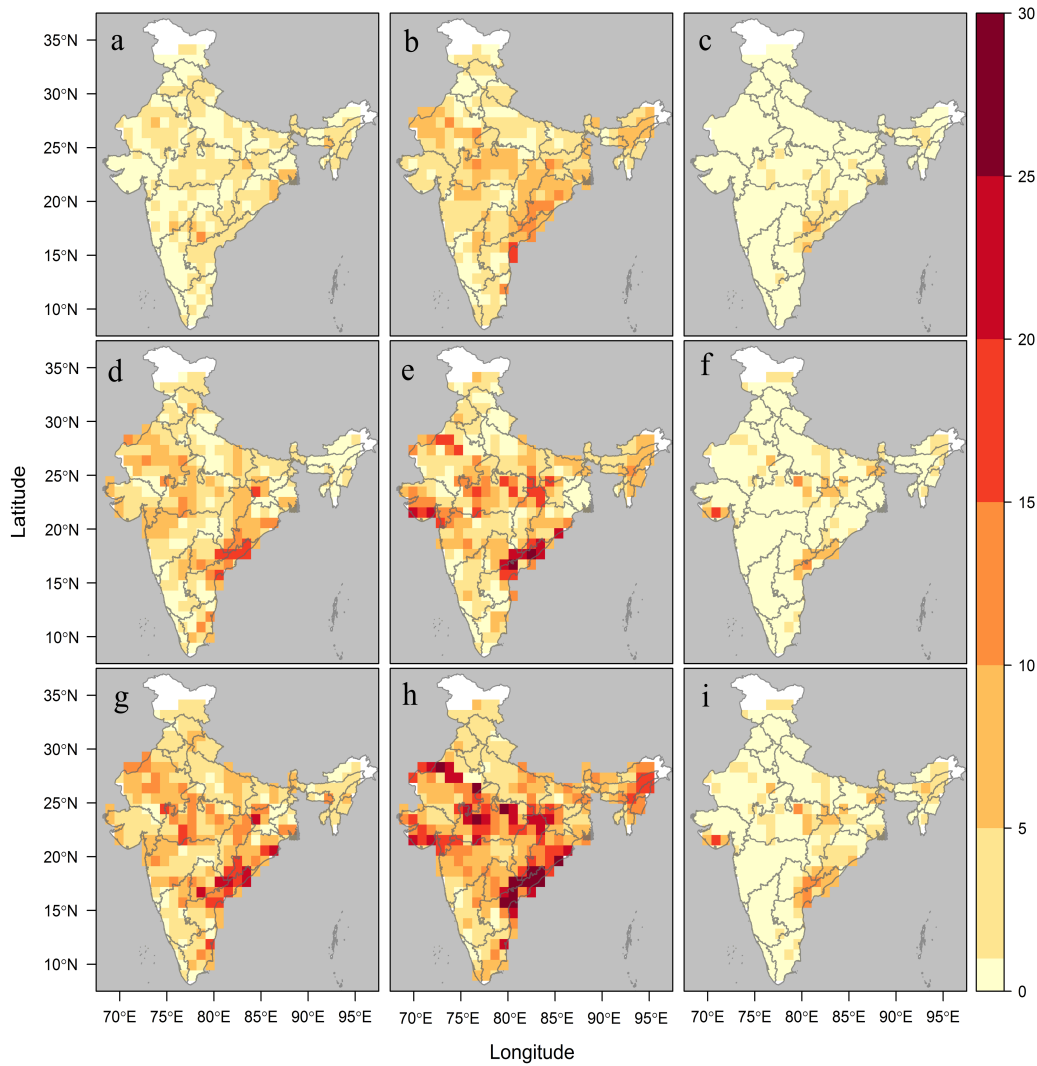

Figure 1. Total number of heatwave events from 1953 - 2013 for different types of heatwaves identified using the EHF (Effective Heat Factor) approach. (a-c) Oppressive heatwaves (days with high temperature and high humidity), (d-f) extreme temperature (days with high temperature and low humidity), (g-i) total heatwave events. (a,d,g) one-day events, (b,e,h) three to five-day events (the condition prevails for three to five consecutive days), (c,f,i) more than five-day events (the condition prevails for more than 5 consecutive days). In a few grids the number of events is more than 30, to show the variability, it is floored to 30 in this figure.

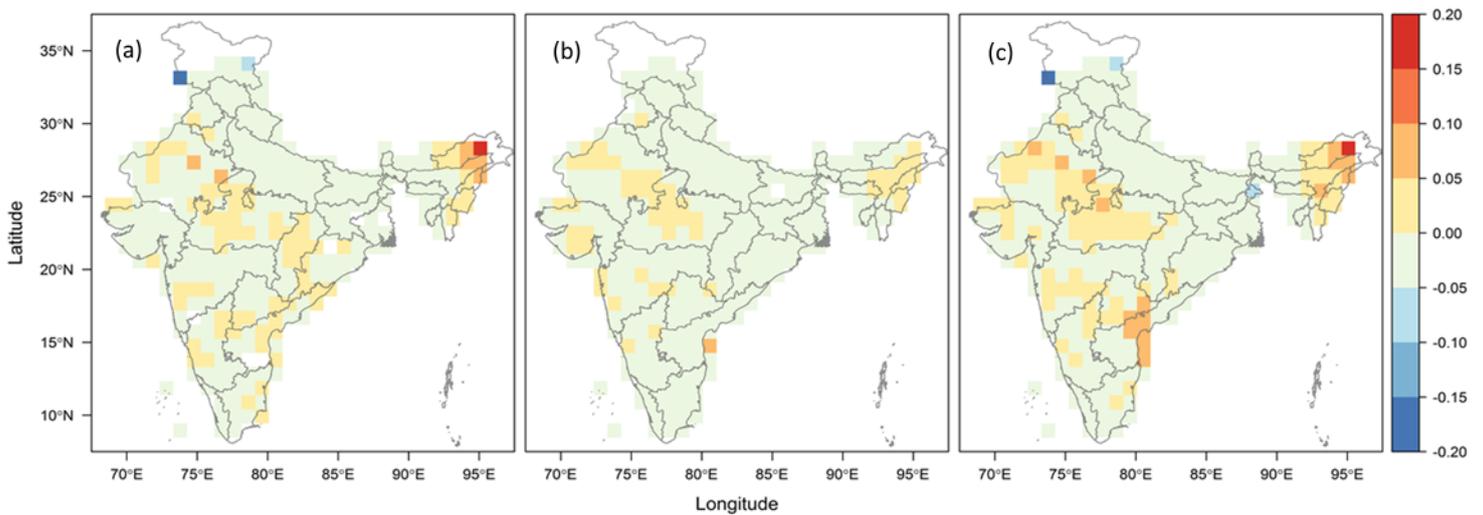

Figure 2. Trend in Heatwave days from 1953 - 2013 in different types of heatwave events (events are not classified based on consecutive days). (a) oppressive heatwaves, (b) extreme heatwaves, and (c) total heatwaves. The trend in each grid is identified using linear regression.

We further examine the association of heat-related mortality with oppressive and extreme heatwave days from 1967 to 2007 (when reliable mortality data is available). The heat-related mortality is strongly positively related to oppressive heatwave days (Pearson correlation coefficient = 0.5) relative to dry heatwave days (Pearson correlation coefficient = 0.17) (Figure 3a & 3c). Consistently, several years (such as 1970, 1972, 1978, 1991, 2003) with spikes in oppressive heatwave days resulted in abnormally high deaths, whereas several years (such as 1976, 1984, 1989, 1997, 1999, 2004) with extreme heatwave days experience lower mortality. These findings are consistent with (Mazdiyasni et al., 2017), which states that mortality due to heatwaves is directly related to the economic conditions of the population. A stable body temperature of approximately ~37°C is maintained in human bodies even when exposed to different environmental conditions (Balmain et al., 2018). Maintaining the body temperature at this narrow, safe range of temperature from 35°C to 39°C is very important to maintain the metabolic reactions at an optimal level (Benzinger, 1969). When thermoregulation fails due to the combination of high temperature and high humidity, the body temperature increases beyond the optimum range, failing multiple systems in the human body. Thus, even when temperature alone, the perspective of heatwaves does not show any alarming conditions induce severe heat stress due to the combination of temperature and humidity. Since oppressive heatwave days significantly increase over India in most climatologically homogeneous zones, we infer the exacerbated heat-related mortality in the future if such heatwave conditions increase monotonically in the future as these have been in recent decades.

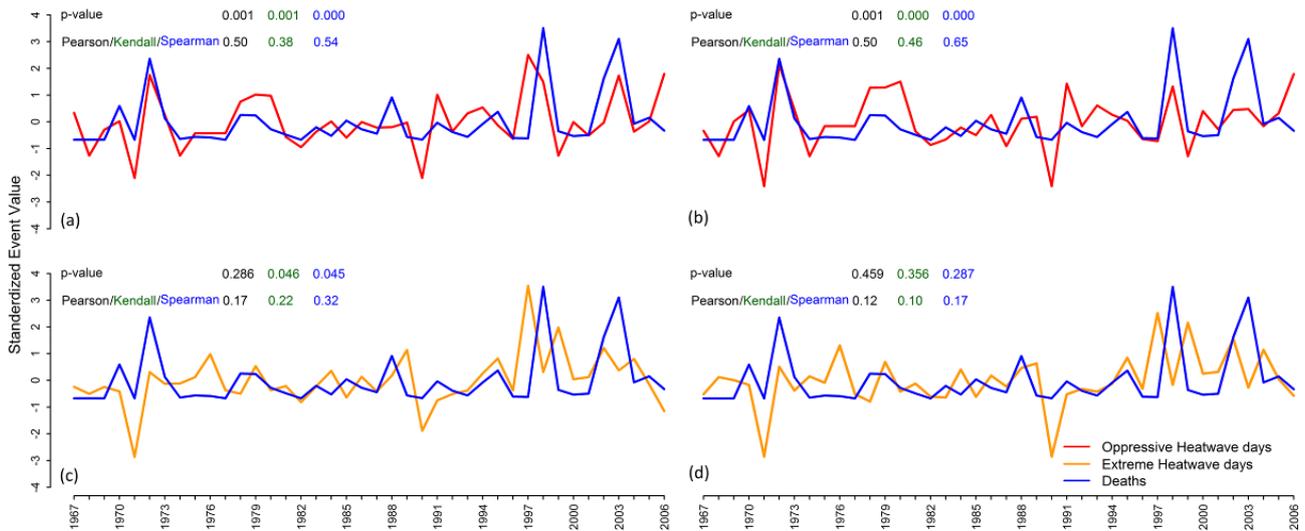

Figure 3. Correlation of Heatwave days (with and without considering acclimatization) with heat-related mortality. (a, b) oppressive heatwave days and (c, d) extreme heatwave days from 1967 – 2006. (a, c) heatwave days without considering acclimatization and (b, d) heatwave days considering at least one-degree sudden rise in temperature.

Further, we examine the role of acclimatization in modulating the relationship between heatwave and heat-related mortality since previous studies suggest that acclimatization significantly suppresses heatwaves' effects on human health (Masdiyazni et al., 2017; Gershunov & Guirguis, 2012; McMichael et al., 2006). Subsequently, to better understand this, we consider an acclimatization factor accounting that the people get acclimatized to the prevailing temperature if the temperature difference between heatwave days and the preceding thirty days is not more than 1°C. Under the acclimatized conditions, heat-related mortality is ~4-fold strongly associated with oppressive heatwave days (Pearson correlation coefficient =0.5) compared to dry heatwave days (Pearson correlation coefficient = 0.12) (Figure 3 c,d). Although the Pearson correlation coefficient between heat-related mortality and oppressive heatwave days is the same in both with and without acclimatized conditions, Kendall's and Spearman's correlation coefficient is higher with acclimatized conditions than without acclimatized conditions (Figure 3 a,b), indicating former shows the actual scenario. Based on these correlations, we infer that oppressive heatwaves are deadlier than extreme heatwaves, and only a sudden change in temperature affects the population adversely, whereas humans get acclimatized if temperature increases gradually (less than 1°C change in two weeks).

*Projected changes in heatwave characteristics in a future warmer climate*

We also aim to quantify future changes in oppressive and dry heatwave days in the mid- (2035-2065) and late-21$^{st}$ (2070-2100) century under 1.5° and 2° warming scenarios. We choose the low warming scenarios as countries ought to restrict the emissions to maintain temperature according to the Paris agreement (2015). We find significant increases (at a 5% significance level) in oppressive heatwave days across Central and Northwestern India in the mid and late 21st century relative to the historical climate (1975-2005) under 1.5° warming. Similarly, we observe an even stronger increase in oppressive heatwave days in both future periods over Central, Western coast, Northwestern, and Northeastern India under 2° warming. In contrast, extreme heatwave days remain unchanged across the country except for the eastern coastal region in both future periods relative to historical under both 1.5° and 2° warming conditions (Figure S5). Aggregately, we find a significant increase (at 5% significance level: p-value ~ 0.00) in the distribution of oppressive heatwave days in the mid and late 21st century relative to historical climate under both 1.5° and 2° warming scenarios (Figure 4). Consistent with observations, dry heatwave days are higher than the oppressive heatwave in the historical period. However, the oppressive heatwave days surpass the extreme heatwave days in both time frames and warming conditions. For instance, the average oppressive heatwave days increase by 5-fold from 3 days in historical to 15 days in both future periods under 1.5° warming.

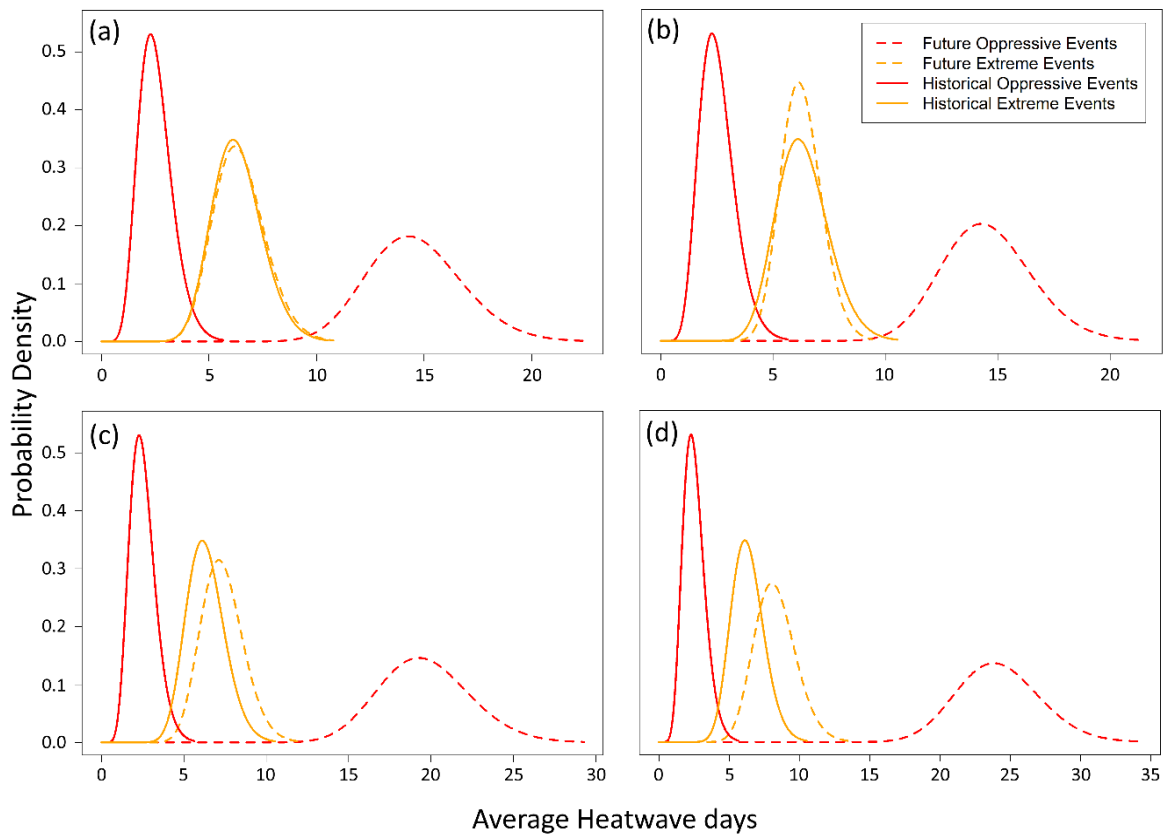

Figure 4. Probabilities of heatwave days in the 1.5° and 2.0° warming scenario. (a) and (b) represent the 1.5° warming scenario, and (c) and (d) represent the 2° warming scenario. In (a) and (c), the period of consideration is 2035 to 2065, and that in (b) and (d) is 2070 to 2100.

Similarly, average oppressive heatwave days increase by ~7-fold in the mid (~21 days) and ~8-fold in the late 21st (~25 days) century relative to historical climate under 2° warming (Figure 5c, d). In contrast, average dry heatwave days remain unchanged under 1.5° and increase slightly under 2° warming in both future periods relative to historical (Figure 4a-d). Restricting the warming to 1.5° from the proposed 2° significantly reduces oppressive heatwave days by 44% in future periods. Given the strong association of heat-related mortality with oppressive heatwaves (Figure 3), such remarkable increases in the oppressive heatwave days present a high potential of elevated heat-related mortality rate in the future amidst inadequate adaptive measures. This humungous increase in oppressive heatwave days can be associated with the increase in average specific humidity over India in the future (Figure 6). The specific humidity remains constant when the temperature rises without a moisture supply (Sherwood & Huber, 2010). Here, as the specific humidity increases with temperature rise, it implies an increase in moisture content in the system. As a result, the $T_E$ increases (a proxy to ambient temperature) with the rise in both temperature and specific humidity, leading to an increase in oppressive heatwaves or converting extreme heatwaves into oppressive heatwaves.

**Conclusions**

This study uses the available heat-related mortality data to examine both the oppressive and extreme heatwave characteristics over India in the historical period, their changes in a future warmer climate, and the effect on human health. The study also examined the effectiveness of the Paris Agreement (2015) in controlling heatwave events in India. We find that the Indian subcontinent experiences a significant rise in the frequency of heatwaves (i.e., the sum of oppressive and extreme heatwaves), consistent with the simultaneous increase in heat extremes across the globe (Perkins et al., 2012; IPCC, 2013), including India (Ratnam et al., 2016; Rohini et al., 2016). In particular, India's average heatwave days per year have increased by ~2.5% in the last six decades. It is further expected to rise until the end of the century under a 2° degree warmer relative to historical climate.

Interestingly, the increasing trend in oppressive heatwave days against no trend in extreme heatwave days indicates that average oppressive heatwave days may surpass the average extreme heatwave days in the warmer future. Specifically, the probability of oppressive heatwave days increases ~8-fold by the end of the 21st century with a 2° warming scenario relative to the historical period. Such a remarkable increase in oppressive heatwave days presents the potential for higher heat-related mortality in the future because historical heat-related mortality is more strongly related to oppressive heatwave days than extreme heatwave days. For instance, the heat-related mortality shows ~3-times (~4-times) higher correlation with oppressive heatwave days with (without) considering acclimatization relative to extreme heatwave days.

The study also gives an insight into the need to keep the carbon emission under check to keep the warming under 1.5°, which reduces the oppressive heatwave days by about 44% relative to 2° warming in future climate. Considering the effect of humidity at 1.5° and 2° warming scenarios, a country like India, which is a highly populated region, could experience heat waves with a magnitude greater than the Russian heatwave of 2010 (the most severe to date) (Lo et al., 2021), which may lead to severe catastrophic events. Additionally, this study suggests that even a moderate and practically unavoidable rise in temperature over the country may lead to an enormous increase in heatwave-related mortality unless the resilience of the vulnerable population is substantially improved. This study mainly concentrated on the number of heatwave days, whereas other aspects like intensity or event duration can be further explored.

**Acknowledgements**

The work presented here is supported by Department of Science & Technology (SPLICE—Climate Change Programme), Government of India (Project reference number DST/CCP/CoE/140/2018, Grant Number: 00000000000010013072 (UC ID: 18192442)).

Supplementary Information

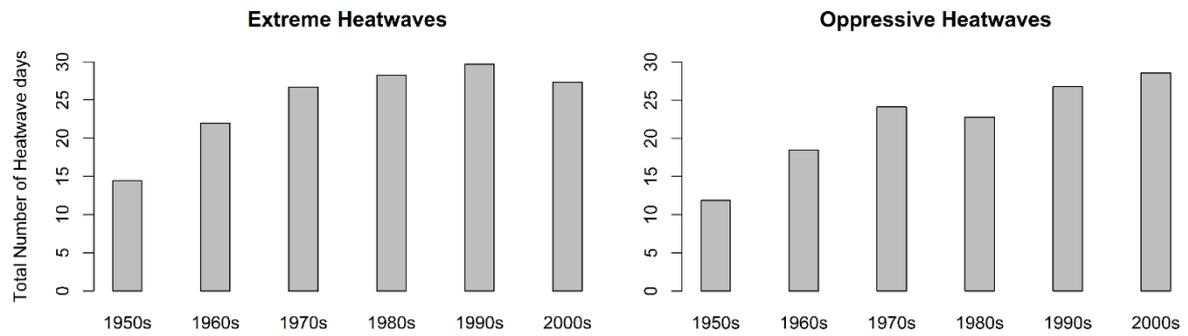

Supplementary Figure S1. Decadal changes in heatwave days. The spatial average of the grids with non-zero values for heatwave days in each year is calculated; from this, the decadal sum is computed.

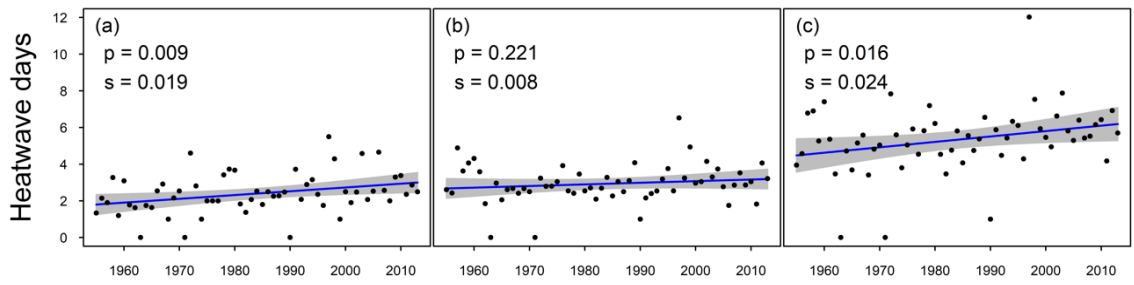

Supplementary Figure S2. The trend in heatwave days. (a) Oppressive heatwave days, (b) Extreme heatwave days and (c) Total heatwave days. The p-value (p) and slope (s) obtained from Theil- Sen Slope estimator are shown in each subplot. The yearly mean of heatwave days is calculated by taking the mean of heatwave days obtained in non-zero grids for the particular year. (0.95 CI is shown in grey colour)

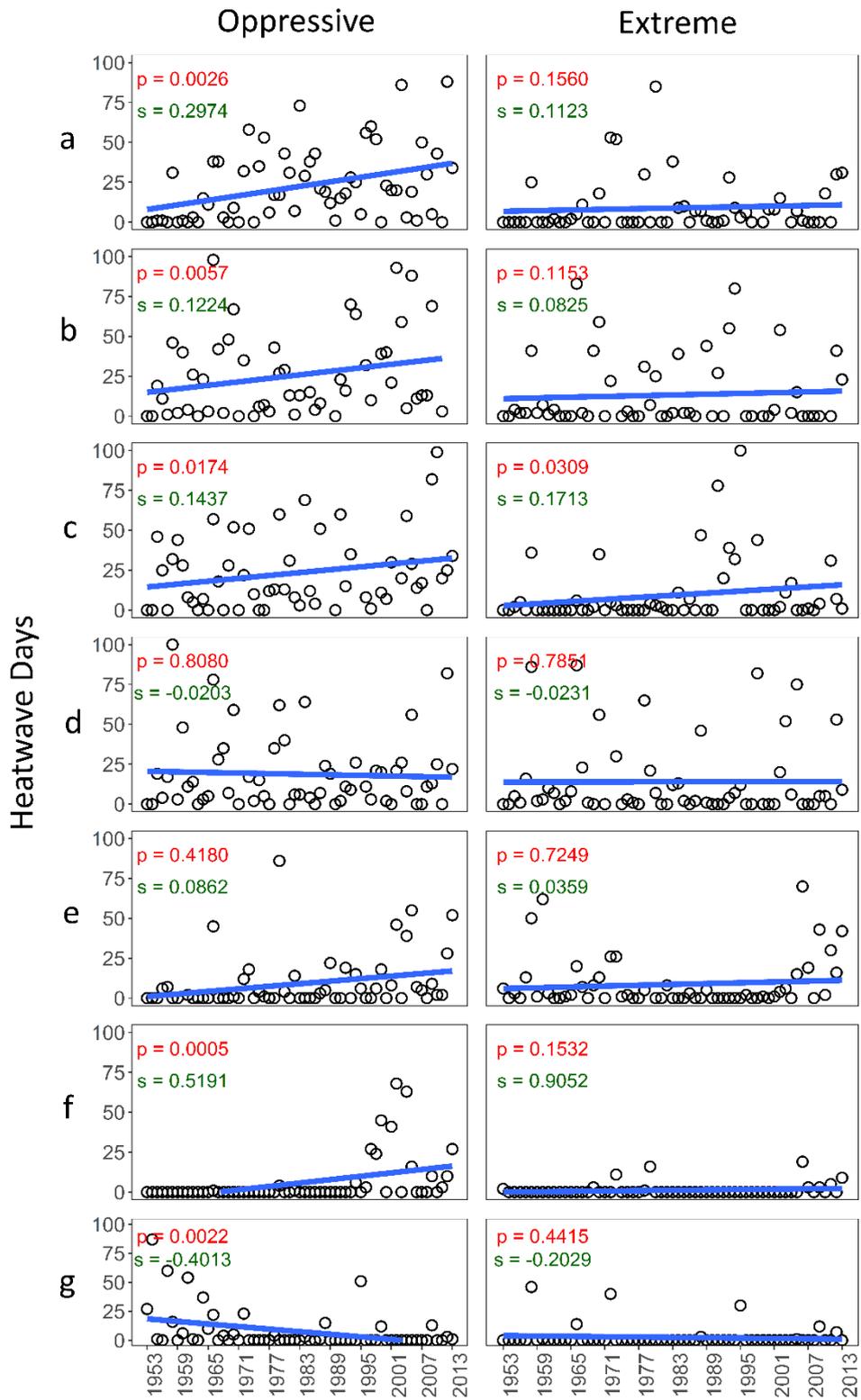

Supplementary Figure S3. Heatwave days trend in each climate zone in India for different types of heatwaves. climate zones: (a) south, (b) central, (c) west, (d) north, (e) north-east, (f) north-east hills and (g) Jammu & Kashmir. The p-value (p) and slope (s) for each sub plot is identified by linear regression. The heatwave days of all grids in a climate zone is totaled in each year to represent the heatwave days in the climate zone.

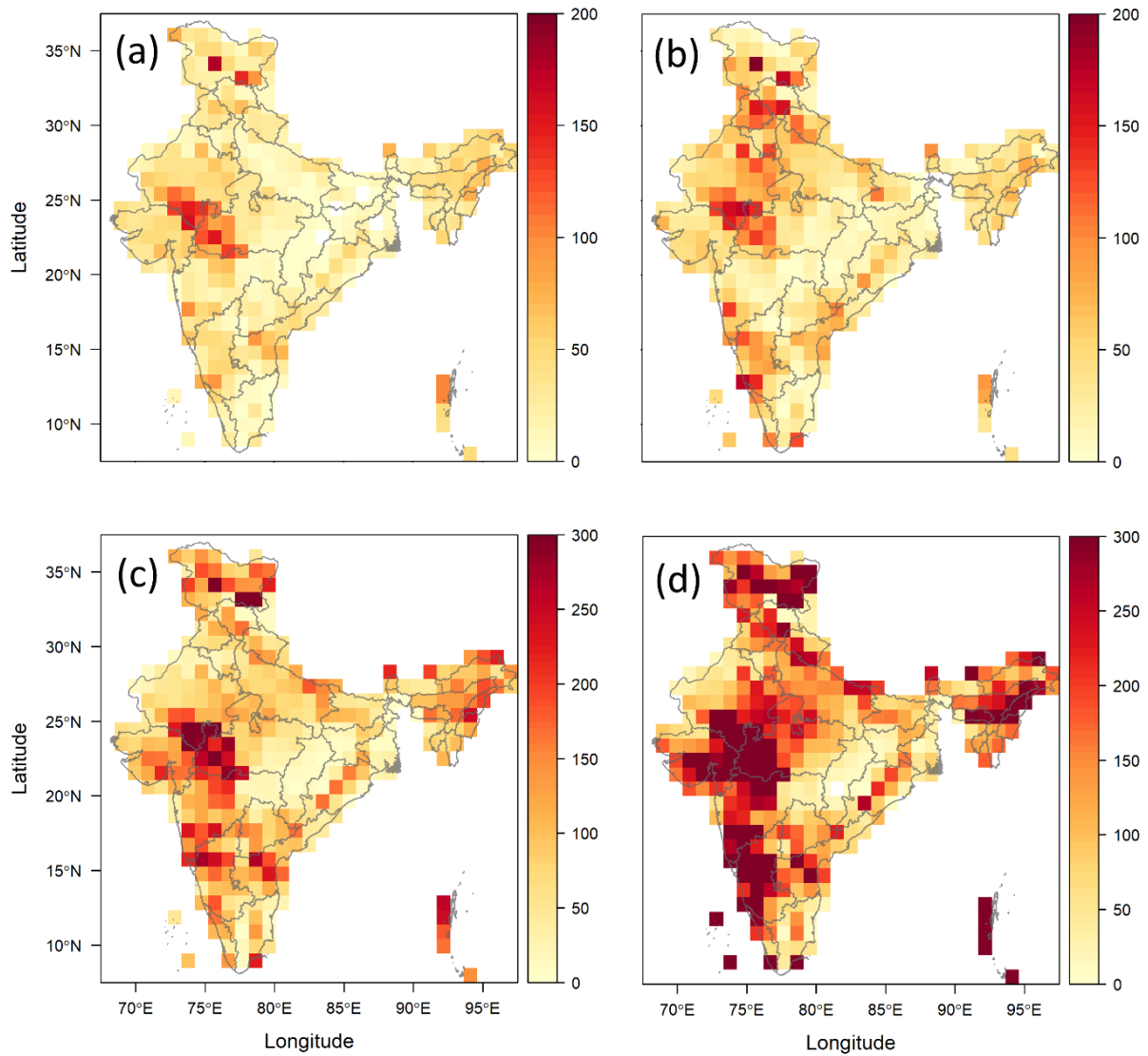

Supplementary Figure S4. Spatial distribution of oppressive heatwave days in low warming conditions. (a) and (b) represents the 1.5° warming scenario and (c) and (d) represent the 2° warming scenario. In (a) and (c), the period of consideration is 2035 to 2065, and that in (b) and (d) is 2070 to 2100

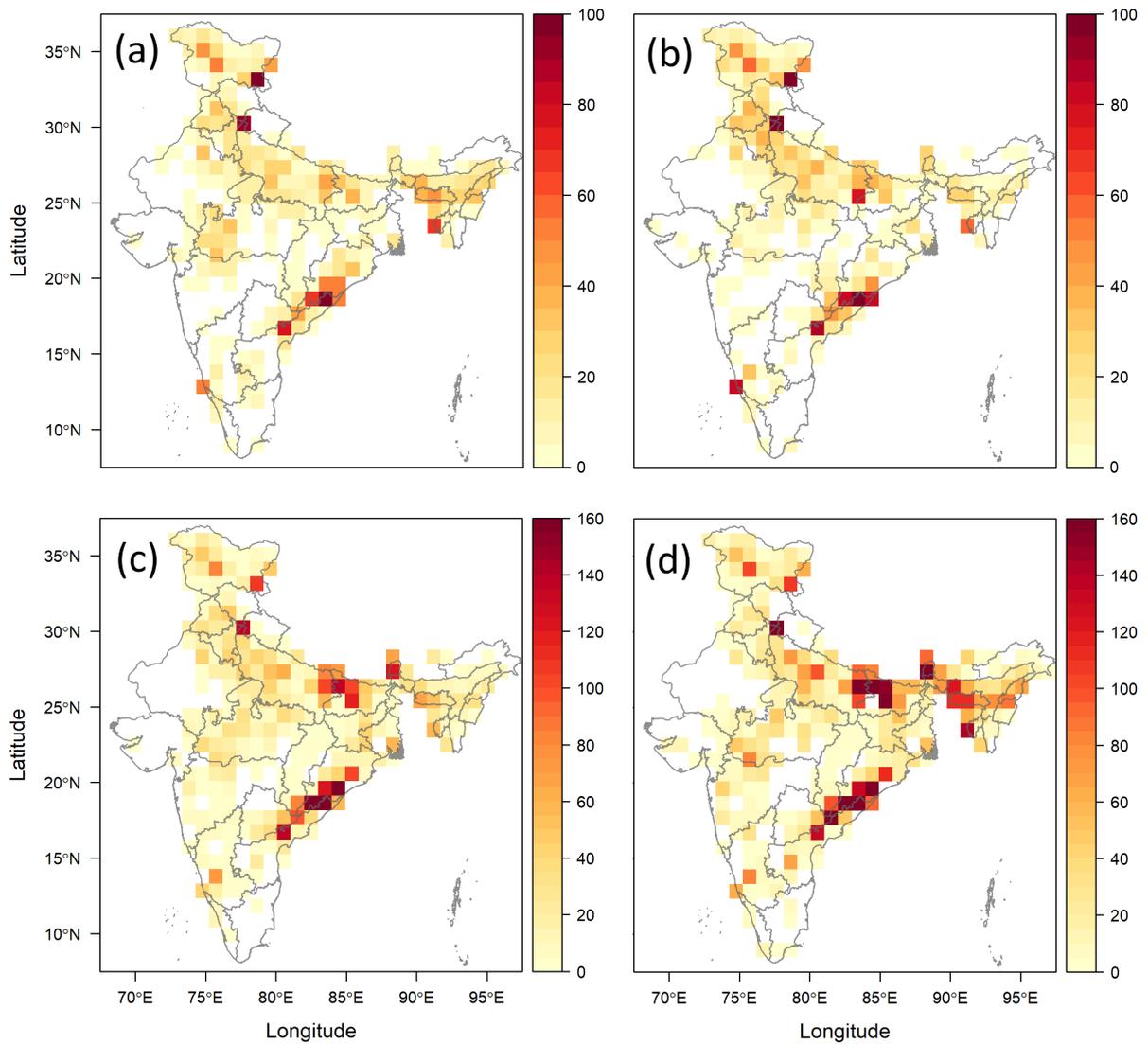

Supplementary Figure S5. Spatial distribution of extreme heatwave days in low warming conditions. (a) and (b) represents the 1.5° warming scenario and (c) and (d) represent the 2° warming scenario. In (a) and (c), the period of consideration is 2035 to 2065, and that in (b) and (d) is 2070 to 2100

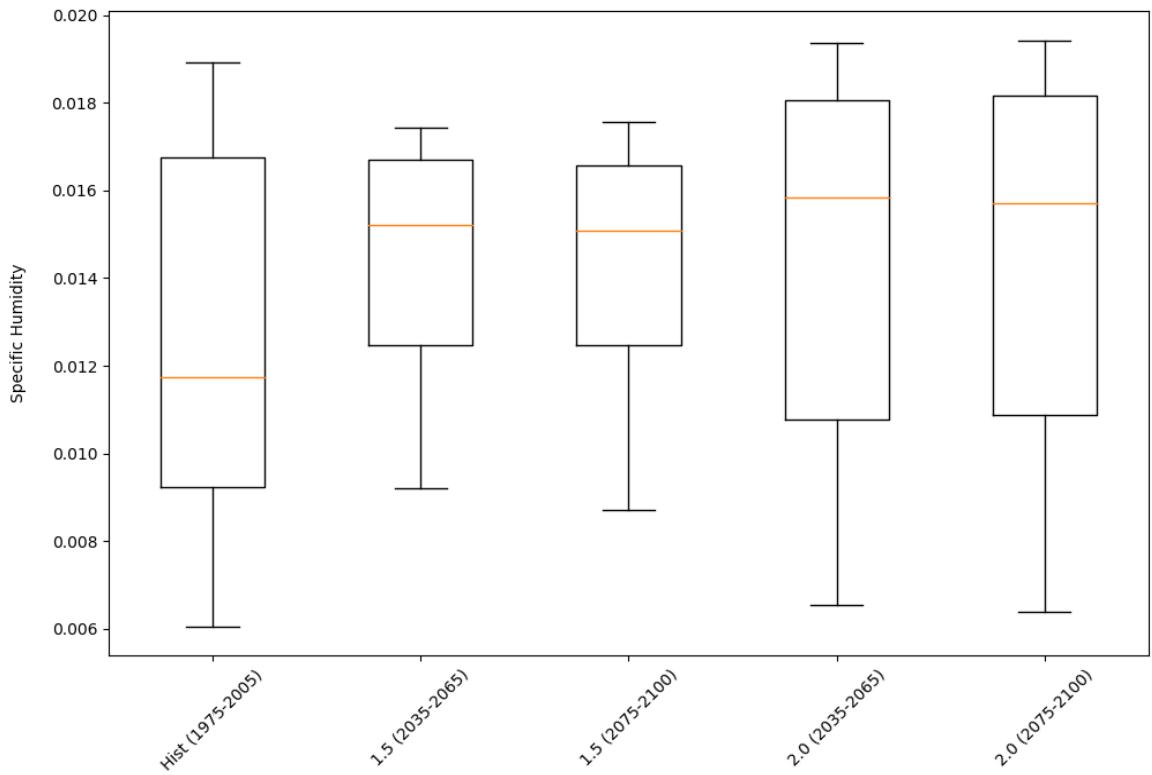

Supplementary Figure S6. Specific humidity during heatwave days in historical and low warming scenarios.

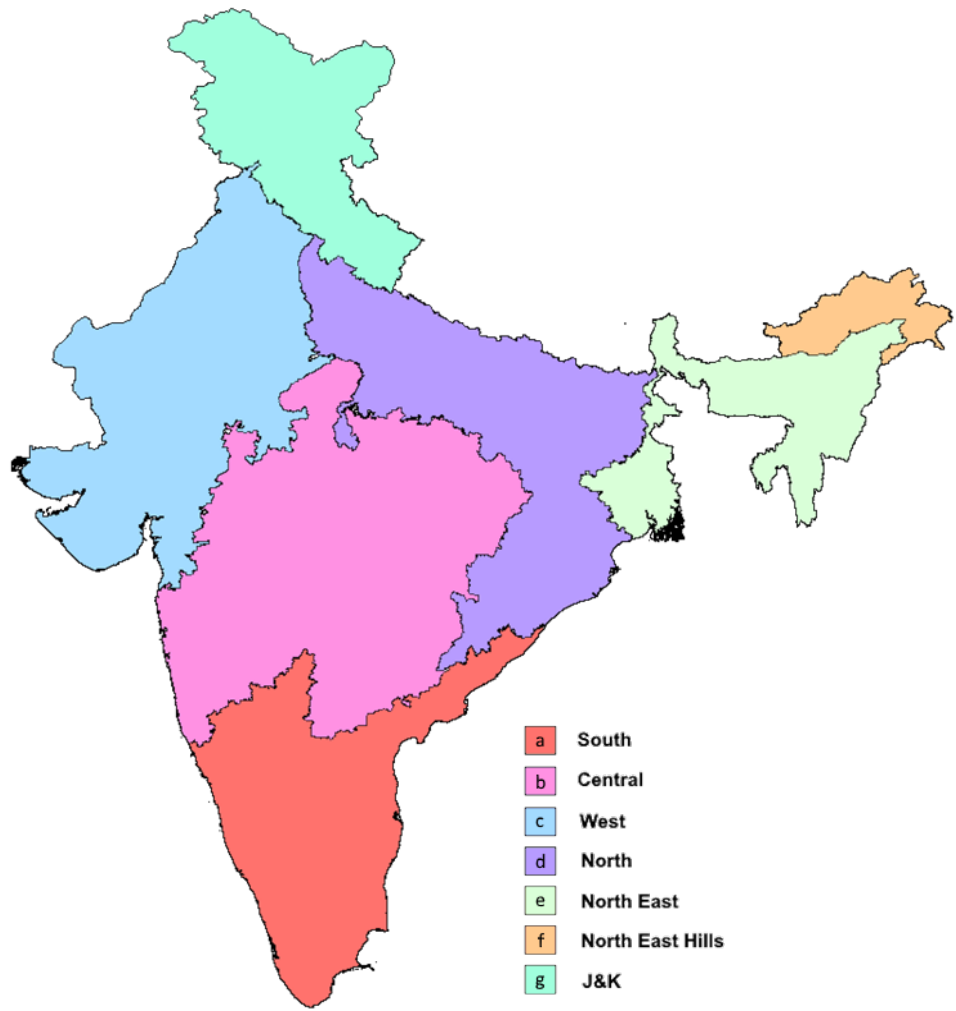

Supplementary Figure S7. Climatologically homogeneous zones